\documentclass[12pt,a4paper]{article}

\usepackage[utf8]{inputenc}
\usepackage{amsmath}
\usepackage{amsfonts}
\usepackage{amssymb}
\usepackage{bbold}
\usepackage[dvips,letterpaper,margin=0.75in,bottom=0.5in]{geometry}
\usepackage{mathrsfs}

\usepackage[T1]{fontenc}
\usepackage[utf8]{inputenc}
\usepackage{authblk}
\usepackage{scrextend}
\usepackage{footmisc}
\usepackage{indentfirst}
\usepackage{physics}
\usepackage[english]{babel}
\usepackage{pdflscape}
\usepackage[noadjust]{cite}
\usepackage{setspace}

\title{Sections and Chapters}
\usepackage{cite}
\title{Inflation and Linear Expansion in the Radiation Dominated era in Jordan-Brans-Dicke Cosmology }
\author{Metin Arik$^{1,}$\footnote{e-mail:metin.arik@boun.edu.tr}}
\author{Medine Ildes$^{1,}$\footnote{e-mail:medine.ildes@boun.edu.tr}}
\author{Mikhail B. Sheftel$^{1,}$\footnote{e-mail:mikhail.sheftel@boun.edu.tr}}
\affil{$^{1}$Department of Physics, Bogazici University, Bebek, Istanbul, Turkiye}

\begin{document}

\maketitle

\begin{abstract}
\indent We present several features of a cosmological model based on the Brans-Dicke-Jordan-Thirry action which is scale invariant with a quartic potential for the Jordan scalar field. We show that the radiation dominated era starts with a closed universe which expands exponentially and the late radiation dominated era expands linearly. We find that there may be a scale-invariant phase with stiff matter between these two radiation dominated eras. The introduction of matter in the linearly expanding universe causes deceleration.
\end{abstract}
\section{Introduction}
\indent The standard cosmology as formulated by Friedman-Le Maitre-Robertson-Walker (FLRW) \cite{friedman1922krummung,friedman1999curvature,lemaitre1927univers,lemaitre1931expansion} has been quite successful in explaining the observed universe. As is well known, in FLRW both radiation and matter cause deceleration of the cosmological expansion. Thus to conform with observations, two eras of accelerated expansion must be assumed. The primordial inflation which immediately follows supposedly radiation dominated big bang and the late acceleration which is usually called the dark energy expressing our hope that the effective cosmological constant observed today can be explained in terms of a related physical phenomenon. \\
\indent In this paper we will consider a cosmological model based on the BDJT Lagrangian density \cite{jordan1938empirischen,jordan1947erweiterung,jordan1959gegenwartigen,thiry1948geometrie,brans1961mach} given by
\begin{align}
\mathcal{L} &=\left[-\Phi R+ \omega\dfrac{1}{\Phi}g^{\mu \nu}\partial_{\mu}{\Phi} \partial_{\nu}{\Phi}-V(\Phi)+L_{M}\right]\sqrt{-g} \\
&=\left[-\frac{1}{8\omega}\phi ^{2}R+\frac{1}{2}g^{\mu \nu}\partial_{\mu}{\phi} \partial_{\nu}{\phi}-V(\phi)+L_{M}\right]\sqrt{-g},
\end{align}
\\where $\phi$ is the Jordan scalar field and $\Phi$ is the Brans-Dicke scalar field, the two being related by
\begin{align}
\Phi=\frac{1}{8\omega}\phi ^{2},
\end{align}
\\where $\omega$ is the dimensionless Brans-Dicke parameter, $R$ is the Ricci scalar and $L_{M}$ represents the contribution due to matter fields. Here we take  the scalar field $\phi$ such that it does not couple to $L_{M}$. We use the metric signature $(+,-,-,-)$ and units with $\hbar=1$, $c=1$. We prefer to use the Jordan definition for the scalar field since in flat spacetime the Lagrangian then becomes the standard Lagrangian for a scalar field
\begin{align}
\mathcal{L} =\frac{1}{2}\partial_{\mu}{\phi} \partial^{\mu}{\phi}-V(\phi).
\end{align}
\\ We will choose the potential $V$ as
\begin{align}
V(\phi)=\frac{1}{4}\lambda\phi ^{4}=16\lambda \omega ^{2} \Phi ^{2},
\end{align}
\\which in flat spacetime leads to a renormalizable quantum field theory. Due to the choice of scale-invariant potential, the coupling constant $\lambda$ is dimensionless, so that there are no dimensional parameters in the BDJT part of the Lagrangian density in (1) and (2).\\
\indent The homogeneous and isotropic cosmological field equations obtained from this Lagrangian density have already been calculated for a potential $V(\phi )=\dfrac{1}{2}m^{2}\phi ^{2}$ by Arik, Calik and Sheftel in \cite{arik2005primordial,arik2006can,arik2008friedmann}. Thus their general form for a potential $V(\phi )$ are given by,
\begin{gather}
\frac{3}{4\omega}\phi ^{2}\left(\frac{\dot{a}^2}{a^2}+\frac{k}{a^{2}}\right)-\frac{1}{2}\dot{\phi }^{2}-V(\phi )+\frac{3}{2\omega}\frac{\dot{a}}{a}\dot{\phi }\phi=\rho, \\
\frac{-1}{4\omega}\phi ^{2}\left(2\frac{\ddot{a}}{a}+\frac{\dot{a}^2}{a^2}+\frac{k}{a^{2}}\right)-\frac{1}{\omega}\frac{\dot{a}}{a}\dot{\phi }\phi-\frac{1}{2\omega}\ddot{\phi }\phi-\left(\frac{1}{2}+\frac{1}{2\omega}\right)\dot{\phi }^{2}+V(\phi )=p,\\
\ddot{\phi }+3\frac{\dot{a}}{a}\dot{\phi }+\dv{V(\phi )}{\phi }-\frac{3}{2\omega }\left(\frac{\ddot{a}}{a}+\frac{\dot{a}^2}{a^2}+\frac{k}{a^{2}}\right)\phi =0,
\end{gather}
\\where $k$ is the curvature parameter with $k=-1,0,1$ corresponding to open, flat, closed universes respectively, $a(t)$ is the scale factor of the universe, dot denotes derivative with respect to time. We will name (6) as the energy density equation, (7) as the pressure equation and (8) as the $\phi$ equation. \\
\indent Sen and Seshadri \cite{sen2003self} have investigated the nature of a potential relevant to the power law expansion in BD cosmology. In \cite{arik2005primordial,arik2006can,arik2008friedmann} a perturbation technique was applied to the above equations with $V(\phi )=\dfrac{1}{2}m^{2}\phi ^{2}$. In addition, exact solutions of modified BD cosmological equations have been found by using symmetry analysis \cite{arik2008symmetry}. On the other hand the quartic potential, $V(\phi )=\dfrac{1}{4}\lambda \phi ^{4}$ has been studied in \cite{mak2002brans} where solutions are comparable with the observed cosmological data only for small negative values of $\omega$ for spatially flat FRW geometry. Chubaryan \textit{et al.} have studied the quartic potential with barotropic equation of state in \cite{chubaryan2009role}. This potential has also been studied in \cite{tahmasebzadeh2017generalized} in the presence of a generalized Brans-Dicke parameter $\omega _{GBD}(\phi)$. Santos and Gregory have found linearly and exponentially expanding solutions for vacuum cosmologies \cite{santos1997cosmology}.\\
\indent Although to investigate role of stiff matter in cosmology was not the main purpose of this work, it appears in our results. An exocit fluid with an equation of state $\nu =p/\rho=1$ was first introduced by Zeldovich \cite{l1962equation}. This fluid is also called as the stiff fluid or Zeldovich fluid and gives energy density proportional to $1/a^{6}$. Many scientists have produced cosmological models with stiff matter \cite{oliveira2011early,dutta2010big,nair2016bulk,chavanis2015cosmology,barrow1978quiescent,joyce1997electroweak,joyce1998turning,colistete2000gaussian,mathew2014cosmology}.
 In addition there exist studies on stiff matter in Brans-Dicke Theory \cite{barrow1986three,kamionkowski1990thermal,lorenz1984exact}. In an another work a complex scalar field description of Bose-Einstein condensate dark matter was studied \cite{li2014cosmological}. It has been found that the early universe evolves from stiff ($p=\rho $) to radiationlike ($p=\rho /3$). We have obtained a similar result in our calculations.
\\ \indent In this paper we will present exact solutions of (6-8) with $V(\phi )=\dfrac{1}{4}\lambda \phi ^{4}$ in a $k=1$ closed universe. We choose $\omega >4\times 10^{4} $ to be confortable compatible with results of Einstein telescope \cite{zhang2017testing} and time delay experiments \cite{alsing2012gravitational}. Our starting point is to introduce a scale invariant solution. Once we find the Jordan scalar field as a function of time, we calculate the scale-factor of the universe for different eras by tracking the behaviour of the field forward and backward in time. Therefore we will have three different $\phi(t)$, we will solve the $\phi$ equation for each case and we will obtain the scale factor as a function of time for each era. These solutions result in inflation in the early radiation dominated era, linear expansion in the late radiation dominated era and scale invariant solution with stiff fluid between these two eras. We also show that introducing matter in the linearly expanding radiation era causes deceleration. Then we calculate temperature-time relations for each era we investigate. Finally we study the passage from big bang to the scale invariant solution and the passage from the scale invariant solution to the linearly expanding solution. 
\section{Radiation Dominated Era}
\subsection{The scale invariant solution}
\indent One important property of potential in (5) is that it does not introduce any dimensional parameters into the Lagrangian density so that the action and the resulting equations are scale invariant. In this part we have assumed that the relation between $\phi (t)$ and $a(t)$ preserves the scale invariance and is given by
\begin{align}
\phi (t)=\frac{A}{a(t)},
\end{align}
\\ where $A$ is a positive dimensionless constant. With this constraint, the $\phi$ equation becomes a second order nonlinear differential equation.
\begin{align}
\frac{A[-3+2A^{2}\omega \lambda-(3+2\omega )\dot{a}^{2}-(3+2\omega )a\ddot{a}]}{2\omega a^{3}}=0,
\end{align}
\\ where dot denotes derivative with respect to time. To be able to solve this equation we introduce new variable $\theta (t)=a^{2}(t)$. Then the differential equation reduces to
\begin{align}
\ddot{\theta}=\frac{4\omega \lambda A^{2}-6}{3+2\omega}.
\end{align}
One can easily find $\theta $ and with appropriate choice of integration constants $b_{1}$ and $b_{2}$ the solution for the scale factor can be written as
\begin{align}
a(t)=\sqrt{\left(\frac{2A^{2}\omega \lambda -3}{3+2\omega}\right)t^{2}+b_{1}t+b_{2}}.
\end{align}
\\Behaviour of energy density and pressure are found to be
\begin{align}
\rho &=A^{2}\left[\frac{6-3A^{2}\omega \lambda}{4a^{4}\omega}+\frac{-(3+2\omega )b_{1}^{2}+4(-3+2A^{2}\omega \lambda)b_{2}}{16a^{6}\omega}\right],\\[2mm]
p&=A^{2}\left[\frac{2-A^{2}\omega \lambda}{4a^{4}\omega}+\frac{-(3+2\omega )b_{1}^{2}+4(-3+2A^{2}\omega \lambda)b_{2}}{16a^{6}\omega}\right].
\end{align}
\\ The term proportional to $1/a^{6}$ is the stiff matter term \cite{l1962equation}. By using the usual continuity equation which is satisfied by (6-8) the terms proportional to $a^{-4}$ are readily recognized as radiation whereas the $a^{-6}$ stiff fluid are related to maximal pressure $p=\rho $  without violation of positivity of energy. Positivity of both terms and real scale factor requires $3/2<A^{2}\omega \lambda <2$  and $b_{2} >(3+2\omega)b_{1}^{2}/4(2A^{2}\omega \lambda -3)$. We should note that constants $b_{1}$ and $b_{2}$ are important and they must not be chosen zero. Solutions before and after this era will be matched by adjusting $b_{1}$ and $b_{2}$. We will call this phase of the universe as the scale invariant phase.\\
\indent As the universe expands the second term becomes negligible Then the equation of state becomes,
\begin{align}
\nu =\frac{p}{\rho}=\frac{1}{3},
\end{align}
\\as it should be in the radiation dominated era. 
\subsection{Linearly Expanding Radiation Dominated Universe}
\indent As time increases its effect in (12) becomes larger so we can make the assumption
\begin{align}
\phi (t)=\frac{B}{t}. 
\end{align}
\\Then $\phi$ equation becomes
\begin{align}
\frac{4\omega Ba^{2}-6\omega Bta\dot{a}+2\omega \lambda B^{3}a^{2}-3Bt^{2}(a\ddot{a}+\dot{a}^{2}+1)}{2\omega t^{3}a^{2}}=0.
\end{align}
\\Firstly we set $a^{2}=\theta$ in the above equation and we obtain
\begin{align}
t^{2}\ddot{\theta}+2\omega t\dot{\theta}-\frac{4\omega}{3}(2+\lambda B^{2})\theta=-2t^{2}.
\end{align}
\\Last equation is easily recognized as non-homogeneous Cauchy-Euler equation and its solution is found as
\begin{gather}
\theta (t)=c_{1}t^{m_{+}}+c_{2}t^{m_{-}}+\frac{3t^{2}}{2\omega(\lambda B^{2}-1)-3}, \\ 
\text{with} \hspace{15pt} m_{\pm}=\frac{1}{2}-\omega \pm \sqrt{\omega ^{2}-\omega +\frac{1}{4} +\frac{4\omega}{3}(2+\lambda B^{2})}, 
\end{gather}
\\where $c_{1}$ and $c_{2}$ are integration constants. Unless one choose $c_{1}=c_{2}=0$, it is impossible to obtain pressure and the density in the form $c/a^{n}$ with constant $c$ and rational $n$. Therefore we obtain the scale factor and the Jordan field as
\begin{align}
a(t)=\sqrt{\frac{3}{2B^{2}\omega \lambda-(2\omega +3)}}t, \hspace{15pt} \phi (t)=\sqrt{\frac{3}{2B^{2}\omega \lambda-(2\omega +3)}}\frac{B}{a(t)}. \\ \nonumber
\end{align}
\\This choice satisfies the $\phi$ equation which always is equal to zero. By using the gravitational field equations one can easily calculate energy, pressure and equation of state as
\begin{align}
\rho &=\frac{9B^{2}[B^{2}\omega \lambda-2(2\omega +3)]}{[2B^{2}\omega \lambda-(2\omega +3)]^{2}}\frac{1}{4\omega a^{4}},\\
p &=\frac{9B^{2}[B^{2}\omega \lambda-2(2\omega +3)]}{[2B^{2}\omega \lambda-(2\omega +3)]^{2}}\frac{1}{12\omega a^{4}},\\
\nu &=\frac{p}{\rho }=\frac{1}{3}. 
\end{align}
\indent Note that in this era although we have not imposed the ansatz $\phi (t)=A^{'}/a(t)$, we have ended up with it. \\
\indent After this point we will continue with results of section 5 where we match solutions for each era. Continuity of $\phi(t)$ and $a(t)$ at the passage from scale invariant phase to linearly expanding era gives 
\begin{align}
a(t)=\sqrt{\frac{2A^{2}\omega \lambda -3}{2\omega +3}}t, \hspace{15pt} B=A\sqrt{\frac{2\omega +3}{2A^{2}\omega \lambda -3}}.
\end{align}
\subsection{Creation of matter in the late radiation dominated era}
\indent We assume that creation of matter in radiation dominated era causes small changes in the Jordan field and in the scale factor. Thus we start the ansatz,
\begin{gather}
\tilde{\phi}(t)=\frac{A}{\mathscr{B}t}+\psi (t) \hspace{15pt} \text{with} \hspace{15pt} \psi (t)=ut^{m}, \\
\tilde{a}(t)=\mathscr{B}t+\alpha (t) \hspace{15pt} \text{with} \hspace{15pt} \alpha (t)=vt^{n}, \\
\mathscr{B}=\sqrt{\frac{2A^{2}\omega \lambda -3}{2\omega +3}}
\end{gather}
\\and $\psi $ and $\alpha $ are small. First we write all three equations in terms of new functions. In addition we neglect second and higher order terms $(\alpha ^{2}, \psi ^{2}, \alpha \psi , ... )$ in the corrections in the perturbations of $\alpha$ and $\psi$. We easily obtain the constant $v\sim t^{2+m-n}$ from the $\phi$ equation. It follows that we must choose $m=n-2$ to keep $v$ constant. Then energy density becomes
\begin{align}
\rho =\frac{C_{1}}{t^{4}}+\frac{C_{2}}{t^{5-n}},
\end{align}
\\with $C_{1}$ and $C_{2}$ are constants. Thus we choose $n=2$ to have two component energy density which is composed of a radiation part and a matter part
\begin{align}
\rho =\frac{C _{r}}{a^{4}}+\frac{C _{m}}{a^{3}},
\end{align}
\\where
\begin{align}
C _{r}&=-\frac{3A^{2}(-2+A^{2}\omega \lambda )}{4\omega }, \\
C _{m}&=-\frac{Au[27+2\omega (6+A^{2}\lambda (-12-5\omega +2A^{2}\omega \lambda (2+\omega)))]}{2\omega [-9-5\omega +2A^{2}\omega \lambda (2+\omega)]}.
\end{align}
\\ Now we will take care of positivity of energy density. We have found $3/2<A^{2}\omega \lambda <2$ at the end of the discussion of section 2.1. In addition $A, \lambda ,\omega $ are all positive parameters. In this scope the parameter $u$ must be chosen as a positive number to ensure positivity of energy density.\\
\indent The parameter which determines whether the universe is accelerating or decelerating is $v$ which is found as
\begin{align}
v=\frac{u(-3+2A^{2}\omega \lambda )[-\omega +2A^{2}\omega \lambda (1+\omega)]}{A(3+2\omega )[-9-5\omega +2A^{2}\omega \lambda (2+\omega)]}.
\end{align}
\\ By using the information given in the last paragraph $v$ is found to be negative. Since $n=2$ (27) tells us that introducing matter in the radiation dominated era causes deceleration.
\section{Early Inflation in the radiation dominated era}
\indent Here we relax our constraint $\phi (t)=\displaystyle\frac{A}{a(t)}$. However we will keep matching our solution for the Jordan field which was found to be
\begin{align}
\phi (t)=\frac{A}{\sqrt{\left(\frac{2A^{2}\omega \lambda -3}{3+2\omega}\right)t^{2}+b_{1}t+b_{2}}}.
\end{align}
\\ We see that as $t$ goes to zero, $\phi$  becomes constant and to investigate this behaviour we look for a solution
\begin{align}
\phi (t)=F.
\end{align}
\\ $\phi$ equation becomes
\begin{align}
F^{3}\lambda -\frac{3F}{2wa^{2}}(1+\dot{a}^{2}+a\ddot{a})=0.
\end{align}
\\We have again used the change of variable method to solve the differential equation by introducing $\theta (t) =a^{2}(t)$. Then the equation immediately reduces to
\begin{align}
\ddot{\theta}-\frac{4\omega \lambda F^{2} \theta}{3}=-2,
\end{align}
and the positive solution for the scale factor of the universe is
\begin{align}
a(t)=\sqrt{c_{1}e ^{\alpha t}+c_{2}e ^{-\alpha t}+\frac{3}{2\omega \lambda F^{2}}}, \hspace{20pt} \alpha =2F\sqrt{\frac{\omega \lambda}{3}}.
\end{align}
\\Energy density and pressure are easily calculated by substitution of $a(t)$ and $\phi (t)$ in equations (6-7)
\begin{gather}
p=\frac{9-16c_{1}c_{2}F^{4}\omega ^{2}\lambda ^{2}}{48\omega ^{2}\lambda a^{4}}, \\[1mm]
\rho =\frac{9-16c_{1}c_{2}F^{4}\omega ^{2}\lambda ^{2}}{16\omega ^{2}\lambda a^{4}}, \\[1mm]
\nu =\frac{p}{\rho }=\frac{1}{3}.
\end{gather}
\\ To have $\rho >0$  we must satisfy the condition $9-16c_{1}c_{2}F^{4}\omega ^{2}\lambda ^{2}> 0$. \\
\indent To simplify our result let us choose $c_{1}=d_{1}/\alpha ^{2}$ and $c_{2}=d_{2}/\alpha ^{2}$ where $d_{1}>0$. Then we have
\begin{align}
a(t)=\frac{1}{\alpha}\sqrt{d_{1}e^{\alpha t}+d_{2}e^{-\alpha t}+2}.
\end{align}
\\ Choosing the integration constant $d_{1}$ and $d_{2}$ such that $a(t)$ is minimum at $t=0$ gives us $d_{1}=d_{2}$ and $a(t)$ must approach a positive non-zero value at $t=0$. Therefore our results become
\begin{align}
a(t)=\frac{\sqrt{2}}{\alpha}\sqrt{d_{1}\cosh(\alpha t)+1}, \\
\rho (t)=\frac{9(1-d_{1}^{2})}{16\omega ^{2}\lambda a^{4}}, \hspace{15pt} p=\frac{\rho}{3},
\end{align}
\\where $0<d_{1}<1$.
\\We can also choose $a(t)=0$ when $t=0$ with $d_{2}=-2-d_{1}$. Then we obtain
\begin{align}
a(t)=\frac{\sqrt{2}}{\alpha}\sqrt{d_{1}\sinh(\alpha t)+1-e^{-\alpha t}},\\
\rho (t)=\frac{9(1+d_{1})^{2}}{16\omega ^{2}\lambda a^{4}}, \hspace{15pt} p=\frac{\rho}{3},
\end{align}
\\with $0<d_{1}$.\\
\indent These results indicate that early inflation took place in the radiation dominated era under the effect of the Brans-Dicke-Jordan field.
\section{Temperature Calculations}
 \indent The relation between energy density, pressure and temperature had already been derived as
\begin{align}
\dv{p(T)}{T}=\dfrac{1}{T}[\rho (T)+p(T)].
\end{align}
\\ In our calculations pressure and energy density is a function of time. Thus we apply chain rule and obtain the temperature as
\begin{align}
T=T_{0}exp\left[\int_{t_{0}}^{t}\dfrac{\displaystyle\dv{p(t^{'})}{t^{'}}}{\rho (t^{'})+p(t^{'})}dt^{'}\right],
\end{align}
\\where $T_{0}$ is the reference temperature of the universe when it evolves with the scale factor $a(t_{0})$. For the early universe, both choices of the scale factor gives the same temperature-time relation which is found as
\begin{gather}
T=T_{0}\frac{a(t_{0})}{a(t)}.
\end{gather}
\\For the scale-invariant phase we have
\begin{align}
T&=T_{0}\frac{\tilde{a}(t)^{2}a(t_{0})^{3}}{\tilde{a}(t_{0})^{2}a(t)^{3}}, \\
\tilde{a}(t)&=\sqrt{\left(\frac{2A^{2}\omega \lambda -3}{3+2\omega}\right)t^{2}+b_{1}t+\frac{(3+2\omega)b_{1}^{2}-4b_{2}}{8(2-A^{2}\omega \lambda)}}.
\end{align}
\\For the linearly expanding radiation dominated era we have obtained
\begin{align}
T=T_{0}\frac{a(t_{0})}{a(t)} \hspace{10pt} \text{or} \hspace{10pt} T=\frac{\tilde{T}_{0}}{t},
\end{align}
\\which is the standard time-temperature relation in the radiation dominated era. \\
\indent For the era where we introduce matter into radiation temperature is more complicated than the other cases and we find the temperature to be given by
\begin{gather}
T=\frac{(c_{1} +uc_{2} t)^{r}}{t} \hspace{15pt} \text{with} \hspace{15pt} \frac{1}{4}<r<1,
\end{gather}
\\where $c_{1}$ and $c_{2} $ are constant. Note that the constant perturbation term $u\ll 1$. Hence by using a series expansion the temperature can be written as
\begin{align}
T=T_{\infty }\left(1+\frac{t_{1}}{t}\right),
\end{align}
\\where $T_{\infty}$ and $t_{1}$ are constant. Physically $T_{\infty}$ will denote the approximate temperature at the end of the radiation dominated era provided that $t_{1}$ can be chosen small compared to the time elapsed from big bang to the end of the radiation dominated era.\\
\section{Matching the solutions}
\indent In this section we will try to match solutions of the scale factor, the Jordan field and the energy density for different eras which follow each other in time. We will not be interested in pressure and temperature because microscopic events can affect them. Let us call the scale factor of the universe $a_{1}(t)$ for the early inflation era, $a_{2}(t)$ for the scale invariant era, and  $a_{3}(t)$ for the late radiation dominated era. Similarly we name the Jordan field solutions as $\phi_{1}(t)$, $\phi_{2}(t)$ and $\phi_{3}(t)$ and the energy densities as $\rho _{1}(t)$, $\rho _{2}(t)$, $\rho _{3}(t)$ respectively. Initially we have tried to match $\phi_{1}(t)$ with $\phi_{2}(t)$ smoothly at a certain time. Then we try to match $a_{1}(t)$ with $a_{2}(t)$ smoothly at the same certain time. Thus we obtain four equations for continuity of $\phi (t)$, $\dot{\phi}(t)$, $a(t)$ and $\dot{a}(t)$. However it is only possible to have three equations satisfied at the same boundary. We eliminate continuity of $\dot{a}(t)$. Consequently we end up with the following equations
\begin{gather}
\phi _{1}(t)=\phi _{2}(t) \hspace{15pt} \text{at} \hspace{5pt} t=t_{1}, \\[1mm]
\dot{\phi}_{1}(t)=\dot{\phi}_{2}(t) \hspace{15pt} \text{at} \hspace{5pt} t=t_{1}, \\[1mm]
a_{1}(t)=a_{2}(t) \hspace{15pt} \text{at} \hspace{5pt} t=t_{1}. 
\end{gather}
\\The passage from early inflation to scale invariant phase occurs at time $t_{1}$ where
\begin{align}
t_{1}=-\frac{3+2\omega}{2(2A^{2}\omega \lambda -3)}b_{1}.
\end{align}
\\When we use $a_{1}(t)=\frac{\sqrt{2}}{\alpha}\sqrt{d_{1}\cosh(\alpha t)+1}$ we obtain
\begin{align}
d_{1}=\frac{2A^{2}\omega \lambda -3}{3\cosh(\alpha t_{1})},
\end{align}
\\so that the condition $ 0<d_{1}<1$ satisfied. When we have $a_{1}(t)=\frac{\sqrt{2}}{\alpha}\sqrt{d_{1}\sinh(\alpha t)+1-e^{-\alpha t}}$
\begin{align}
d_{1}=\frac{2A^{2}\omega \lambda -3}{3\sinh(\alpha t_{1})}+\frac{e^{-\alpha t_{1}}}{\sinh(\alpha t_{1})},
\end{align}
\\so that the condition $0<d_{1}$ satisfied.\\
\indent By using the information obtained above and $3/2<A^{2}\omega \lambda<2$ one can easily compare energy densities for the early universe and the scale-invariant phase. It is seen that there is a loss in energy density at the passage.\\
\indent Now we will study matching the scale-invariant phase with the linearly expanding radiation dominated era. We can satisfy only continuity of $a(t)$ and $\phi (t)$. This gives the following results
\begin{align}
t_{2}=\frac{-b_{2}}{b_{1}}, \hspace{15pt} B=A\sqrt{\frac{2\omega +3}{2A^{2}\omega \lambda -3}}.
\end{align}
\\Therefore when we combine the outcomes of the two matching procedure, scale factors for second and third eras can be written as
\begin{align}
a_{2}(t)&=\sqrt{(\frac{2A^{2}\omega \lambda -3}{2\omega +3})[t^{2}+2t_{1}(t_{2}-t)]}, \\
a_{3}(t)&=\sqrt{\frac{2A^{2}\omega \lambda -3}{2\omega +3}}t. 
\end{align}
\\Again it is found that there is a loss in the energy density at the passage.
\section{Conclusion}
\indent There are three gravitational field equations given by (6-8) with four unknowns $a(t)$, $\phi (t)$, $\rho (t)$ and  $p(t)$. Therefore it is impossible to solve them without any further information. Usually one chooses an appropriate energy density to find the solution for the desired era. However our approach to the problem is different. We have used the scale invariant ansatz $\phi(t)=\displaystyle A/a(t)$ and obtained an exact solution for the Jordan field and the scale factor which evolves from radiation and stiff fluid combined phase to a radiation phase. Similar result have been found in the standart model \cite{li2014cosmological} before our study. We have named this combined phase as the scale invariant phase and by investigating the behaviour of the Jordan field backward in time, we have found a universe which starts to expand exponentially at big bang with pure radiation. Similarly, investigating the behaviour of the Jordan field by extrapolating forward in time we again obtain a pure radiation dominated phase which expands linearly. We have found that introducing matter in this linearly expanding late radiation era causes deceleration. Furthermore  we have presented the time-temperature relations for each era. As a result we have not only found the scale factor and the Jordan field for each era but also we have found the order of the relevant eras in time. In the last part of the calculations we have matched solutions for $\phi(t)$ and $a(t)$ at the boundaries of the eras. Thus the three important features of closed space-like section, radiation domination and primordial inflation can be explained by JBDT model. \\ 
\section*{Acknowledgement}
\indent We would like to acknowledge fruitful discussion about Zeldovich fluid with Nihan Katirci. We thank Bogazici University for the financial support provided by the Scientific Research Fund (BAP), research project No 11643.

\bibliographystyle{unsrt}
\bibliography{phi_to_4}
\end{document}